\documentclass[onecolumn,floatfix,showpacs,aps,nofootinbib]{revtex4}

\usepackage[dvips]{epsfig}
\usepackage[english]{babel}

\usepackage{bbm}
\usepackage{verbatim}
\usepackage{array}
\usepackage{amsmath}
\usepackage{multirow}
\usepackage{amsfonts}
\usepackage{amssymb}
\usepackage{xcolor}
\usepackage{hyperref}
\usepackage{leading}
\hypersetup{colorlinks=true,linkbordercolor=blue,linkcolor=blue, citecolor=blue}
\usepackage{subeqnarray}
\usepackage{setspace}
\usepackage{indentfirst} 

\begin{document}

\title{Type-4 spinors: transmuting from Elko to single-helicity spinors}

\author{C. H. Coronado Villalobos$^{1}$} \email{carlos.coronado@inpe.br}
\author{R. J. Bueno Rogerio$^{2,3}$} \email{rodolforogerio@feg.unesp.br}
\author{E. F. T. S\~ao Sabbas$^{1}$} \email{eliahfersaosabbas@gmail.com}
\affiliation{$^1$Instituto Nacional de Pesquisas Espaciais (INPE),\\
12227-010, S\~ao Jos\'e dos Campos, SP, Brazil. \\$^2$Universidade Estadual Paulista (UNESP)\\Faculdade de Engenharia, Guaratinguet\'a, Departamento de F\'isica e Qu\'imica\\
12516-410, Guaratinguet\'a, SP, Brazil}
\affiliation{$^{3}$Instituto de F\'isica e Qu\'imica, Universidade Federal de Itajub\'a - IFQ/UNIFEI, \\
Av. BPS 1303, CEP 37500-903, Itajub\'a - MG, Brasil.}


\begin{abstract}

{\textbf{Abstract.}}
In this communication we briefly report an unexpected theoretical discovery which emerge from the mapping of Elko mass-dimension-one spinors into single helicity spinors. Such procedure unveils a class of spinor which is classified as  type-4 spinor field within Lounesto classification. In this paper we explore the underlying physical and mathematical contents of the type-4 spinor.

\end{abstract}


\maketitle

\section{Introduction}
According to the Lounesto classification there are six disjoint classes of spinors \cite{lounestolivro}. The first three classes stand for regular spinors fields for spin $1/2$ fermions and the remaining three classes stand for singular spinors like type-4, Majorana and Weyl spinors fields.
Until the present moment, all the above mentioned spinors (and the associated quantum fields) are well-established except for the type-4 spinor. Since it was proposed by Lounesto \cite{lounestolivro}, no physical or experimental evidence was found, therefore, we do not know what kind of particle is described by the mentioned class of spinors. 
The classical path that several physicists and mathematicians normally use to construct such spinors is based on the so-called \emph{inversion theorem} \cite{crawford1, crawford2}, however, the last mentioned protocol do not explicitly provide the spinor's components. In other words, it only shows how components might be connected, therefore, we do not have enough information about how it emerges from the space-time symmetries. It does not bring information about the related dynamics and the behaviour under certain discrete symmetries like parity ($P$), charge conjugation ($C$) and time-reversal ($T$). Consequently, all the above mentioned features remain open for investigation. The focus of the present communication is to find explicitly a spinor to fill this gap in the Lounesto classification and shed some light on several related areas.
 
Although there is no quantum field operator constructed based on type-4 spinors, it is expected that it does not respect the full Lorentz symmetries. This spinors are potential candidates to describe dark matter, dark energy and to construct mass-dimension-one fermions \cite{cavalcanti4}. While there is compelling evidence in astrophysics and cosmology that most of the mass of the Universe is composed of a new form of non baryonic dark matter, there is a lack of evidence of the existence of new physics at LHC (Large Hadron Collider) and other particle physics experiments. On the theory side, many specific models with new particles and interactions beyond the standard model have been proposed to account for dark matter \cite{mirco}. The type-4 spinors fields are wealthy regarding their mathematical structure \cite{lounestolivro, da2010elko} and from the Physics point of view have been found to be they particle corresponding to the solution of the Dirac equation in $f(R)$ gravity with torsion \cite{esk}. 

In this paper we report an unexpected theoretical discovery which comes from mapping of Mass-Dimension-One (MDO) spinors into single helicity spinors. MDO spinors were firstly introduced in the literature around the 90's. Elko spinor fields compose a complete set of dual helicity spinor that are neutral under charge conjugation operator. Because of this, Elko fields have suppressed interactions with the Standard Model particles. In such a way, these fields are dark with respect to the matter and gauge fields of the Standard Model, interacting only with gravity and the Higgs boson \cite{hoffdirac}. 
The net result of such mapping protocol surprisingly enough give rise to the type-4 spinors and also to regular spinors (it can be accomplished by fixing an arbitrary phase parameter). This given protocol must be understood as an attempt to incorporate dark matter to the Standard Model. The present paper give allowance to transmute from the MDO field to another kind of MDO field and also from MDO field to Dirac field. Evidently given task was obscure until the present days because  type-4 spinors ``born'' from MDO spinors. Therefore, only Elko spinors were considered to belong to a class of mass-dimension-one fields, however, type-4 spinors carry the same mass-dimensionality as Elko's do, in the meantime, they are single helicity spinors. Interestingly enough, it is the first time that we observe a single helicity   spinor endowed with mass-dimension-one feature. These new objects endowed with mass-dimension-one are strong candidates to describe dark matter and perhaps dark energy. The difference among type-4 spinors and Elko spinors lies on the fact that the first type have no suppressed interaction with the Standard Model of particle physics. So, the last statement translates into the fact that it does not exhibit neutral character under charge conjugation operator. 

In the present work we explore the underlying details related to the construction of the type-4 spinors and their physical information. Here we report what we have deciphered about the associated dynamics, spin sums calculation and the encoded physical information, looking towards a possible quantization. The paper is organized as follows: Section \ref{sec2} is an in-depth overview about MDO and single helicity spinors, highlighting the main aspects of both spinors. In the Sec \ref{sec3} we define and establish the mapping protocol employed throughout this work. In the Section \ref{sec4} we explore the physical information, i.e., we evaluate the bilinear covariants. We reserved Section \ref{sec5} to study the dynamics associated with type-4 spinors and compute the spin sums, based on the new observed aspects inherited from Elko spinors. Finally, Section \ref{sec6} we conclude.

\section{Elementary overview on mass-dimension-one and single-helicity spinors}\label{sec2}
Since they were proposed, the Dirac spinors are well known to describe a specific particle: the electron. Most text-books spend dozen of pages defining the Dirac spinor main features, e.g., the related dynamics, the quantum field operator, bilinear structures, interactions and couplings,  \emph{etc}. Dirac spinors emerge naturally from the symmetries of the full Poincar\'e group\footnote{By full Poincar\'e group we mean boosts, rotations, space-time translations, parity and time-reversal symmetries.}, and the particle interpretation depends on some given properties under transformations by certain group transformations \cite{Wigner1}. 

Dirac spinors are defined as a single helicity objects, whose representation spaces are related by the parity symmetry. In this sense, they compose a complete set of eigenspinors of the parity operator and are governed by the Dirac dynamics (for more details, please, check the References \cite{ryder, nondirac, diracpauli}). So, in the Weyl representation, Dirac spinors are defined as   
\begin{equation}\label{diracspinor}
\psi_{D}(\boldsymbol{p}) = \left(\begin{array}{c}
 \phi_R (\boldsymbol{p}) \\ 
\phi_L (\boldsymbol{p})  
\end{array}\right),
\end{equation}
commonly $\phi_R (\boldsymbol{p})$ and $\phi_L (\boldsymbol{p})$ are defined as right- and left-hand components under Lorentz transformations, respectively. Recently, a set of dual helicity spinors which compose a complete set of eigenspinors of the charge conjugation operator, namelly Elko (or mass-dimension-one spinors), were reported in the literature \cite{jcap, aaca}. These spinors only respect the Klein-Gordon wave-equation and due to this, their fields are endowed with mass-dimension-one \cite{jcap, aaca}. Elko spinors show some peculiar, and very particular, properties that make them interesting objects to explore. Elko spinors are built in the absence of discrete symmetries as a fundamental relation between the representation space. They are defined within proper orthochronous Lorentz subgroup $(\mathcal{L}_{+}^{\uparrow})$ \cite{bilineares}. Nevertheless, Elko's representation space are connected by the Wigner's spin $1/2$ time reversal operator
\begin{equation}
\Theta=\left(\begin{array}{cc}
0 & -1 \\ 
1 & 0
\end{array}\right), 
\end{equation}
which has the property $\Theta\sigma\Theta^{-1}=-\sigma^*$. 
Therefore, the Elko spinors are then defined as
\begin{equation}\label{elkospinor}
\lambda^{S/A}_{h}(\boldsymbol{p}) = \left(\begin{array}{c}
\pm i\Theta[\phi_L(\boldsymbol{p})]^* \\ 
\phi_L(\boldsymbol{p})
\end{array}\right), 
\end{equation}
where $\Theta[\phi_L(\boldsymbol{p})]^*$ and $\phi_L(\boldsymbol{p})$ are defined as right- and left-hand components, the upper indexs $S$ and $A$ stand for self-conjugated and anti-self-conjugated under charge conjugation operation ($\mathcal{C}\lambda^{S/A}_{h}=\pm\lambda^{S/A}_{h}$) and the lower index $h$ stands for the helicity of each component \cite{aaca}.

By dual helicity feature we mean once the left hand helicity is stablished, e.g.,
\begin{eqnarray}\label{helicidadeleft}
\vec{\sigma}.\hat{p}\;\phi_L^{\pm}(\boldsymbol{p})=\pm\phi_L^{\pm}(\boldsymbol{p}),
\end{eqnarray}
the other component has opposite helicity, i.e.,
\begin{equation}\label{helicidaderight}
\vec{\sigma}.\hat{p}\;\Theta[\phi_L^{\pm}(\boldsymbol{p})]^* = \mp \Theta[\phi_L^{\pm}(\boldsymbol{p})]^*,
\end{equation}
contrasting with the single helicity case \cite{nondirac} 
\begin{equation}
\vec{\sigma}.\hat{p}\;\phi_{R/L}^{\pm}(\boldsymbol{p})=\pm\phi_{R/L}^{\pm}(\boldsymbol{p}),
\end{equation}
where the helicity operator reads
\begin{equation}\label{matrixhelicity}
\vec{\sigma}.\hat{p}=\left(\begin{array}{cc}
\cos(\theta) & \sin(\theta)e^{-i\phi} \\ 
\sin(\theta)e^{i\phi} & -\cos(\theta)
\end{array} \right),
\end{equation}
representing the unit vector along $\boldsymbol{p}$ as $\hat{p}= (\sin(\theta)\cos(\phi), \sin(\theta)\sin(\phi), \cos(\theta))$.
Note that the dual helicity feature comes from the Wigner time-reversal operator. So, Elko spinors are pioneer objects in the literature to carry dual helicity feature.

\section{Mapping Elko spinors into single helicity spinors: The rise of a hidden class}\label{sec3}
In the previous section we discussed about the Elko and the single helicity spinors fundamental details.  A natural question that may rise is: Is it possible, through some mathematical treatment, to transmute Elko's dual helicity and make it become a single helicity object as Dirac is do? Does this procedure bring any relevant physical and mathematical information? To answer these questions, we start with the following assumption: 
\begin{eqnarray}\label{usual}
\psi(\boldsymbol{p})=\mathcal{M}\lambda^{S/A}(\boldsymbol{p}),\label{map}
\end{eqnarray}
where $\mathcal{M}$ is a $4\times 4$ mapping matrix that extinguishes Elko dual helicity feature. 
To find the $\mathcal{M}$ components, we manipulate mathematically the relations \eqref{helicidadeleft} and \eqref{helicidaderight}, imposing the following condition to the $\lambda^{S}_{\{-+\}}(\boldsymbol{p})$ spinors\footnote{A very same mathematical treatment is also valid for the $\lambda^{A}$ spinors \emph{mutatis mutandis}}
\begin{eqnarray}
\vec{\sigma}.\hat{p}\;\mathcal{M}_{1_{(11)}}\Theta[\phi_L^{+}(\boldsymbol{0})]^*=+\mathcal{M}_{1_{(11)}}\Theta[\phi_L^{+}(\boldsymbol{0})]^*,\label{Ian}
\end{eqnarray}
where
\begin{eqnarray}
\mathcal{M}_{1}=\left(\begin{array}{cc}
\mathcal{M}_{1_{(11)}} & \mathcal{M}_{1_{(12)}}\\
\mathcal{M}_{1_{(21)}} & \mathcal{M}_{1_{(22)}}
\end{array}
\right).
\end{eqnarray}
Please note that $\mathcal{M}_{1_{(ij)}}$ is a $2\times 2$ matrix. Nevertheless, if one wishes to impose the last condition, given in \eqref{Ian}, to the component that transforms as left-hand component, its helicity should remain unchanged. For this reason, the last statement translates into $\mathcal{M}_{1_{(22)}}=\mathbbm{1}$ and $\mathcal{M}_{1_{(12)}}=\mathcal{M}_{1_{(21)}}=0$. Then, after a straightforward calculation, one gets the following matrix
\begin{eqnarray}
\mathcal{M}_{1}(\boldsymbol{p}) =  \left(\begin{array}{cccc}
m_{11} & (\kappa_{1}+m_{11}\tan(\theta/2))e^{-i\phi} & 0 & 0 \\ 
(-\kappa_{1}+m_{22}\cot(\theta/2))e^{i\phi} & m_{22} & 0 & 0 \\ 
0 & 0 & 1 & 0 \\ 
0 & 0 & 0 & 1
\end{array}  \right),\label{Lucca}
\end{eqnarray}
where $\kappa_1$ is an arbitrary parameter, its fixation allow us to obtain regular or singular spinors. So, the structure of the mapped single helicity spinor, namelly $\psi(\boldsymbol{p})$ read \footnote{To reduce the notation we have defined the boost parameters as $\mathcal{B}^{\pm}(p)\equiv \sqrt{\frac{E+m}{2m}}\big(1\pm \frac{p}{E+m}\big)$}
\begin{eqnarray}\label{psi1}
\psi_{1\uparrow}(\boldsymbol{p})=\sqrt{m}\left(\begin{array}{c}
i\kappa_1\mathcal{B}^{+}(p)\cos(\theta/2)e^{-i\phi/2}\\
i\kappa_1\mathcal{B}^{+}(p)\sin(\theta/2)e^{i\phi/2}\\
\mathcal{B}^{-}(p)\cos(\theta/2)e^{-i\phi/2}\\
\mathcal{B}^{-}(p)\sin(\theta/2)e^{i\phi/2}\\
\end{array}
\right).
\end{eqnarray}
Note that the employed mapping procedure only changes the helicity of the component that transforms like a right-hand component\footnote{The symbols $\uparrow$ and $\downarrow$ stand for the positive and negative helicity of the components.}. However, it is possible to keep the right-hand helicity and then change the helicity of the left-hand component, performing the following transformation
\begin{eqnarray}
\vec{\sigma}.\hat{p}\;\mathcal{M}_{2_{(22)}}\phi_L^{+}(\boldsymbol{0})=-\mathcal{M}_{2_{(22)}}\phi_L^{+}(\boldsymbol{0}).
\end{eqnarray}
From the above condition, we are able to write the next mapping matrix 
\begin{eqnarray}
\mathcal{M}_{2}(\boldsymbol{p}) =  \left(\begin{array}{cccc}
1 & 0 & 0 & 0 \\ 
0 & 1 & 0 & 0 \\ 
0 & 0 & m_{34} &-(\kappa_{2}+m_{33}\cot(\theta/2))e^{-i\phi}  \\ 
0 & 0 & (\kappa_{2}-m_{44}\tan(\theta/2))e^{i\phi} & m_{44}
\end{array}  \right),
\end{eqnarray}
resulting in
\begin{eqnarray}\label{psi2}
\psi_{2\downarrow}(\boldsymbol{p})=\sqrt{m}\left(\begin{array}{c}
-i\mathcal{B}^{+}\boldsymbol(p)\sin(\theta/2)e^{-i\phi/2}\\
i\mathcal{B}^{+}\boldsymbol(p)\cos(\theta/2)e^{i\phi/2}\\
-\kappa_{2}\mathcal{B}^{-}\boldsymbol(p)\sin(\theta/2)e^{-i\phi/2}\\
\kappa_{2}\mathcal{B}^{-}\boldsymbol(p)\cos(\theta/2)e^{i\phi/2}
\end{array}
\right).
\end{eqnarray} 
Now, with this protocol at hand, we are able to map the  $\lambda^{S}_{\{+-\}}(\boldsymbol{p})$ spinor
\begin{eqnarray}
\vec{\sigma}.\hat{p}\;\mathcal{M}_{3}\Theta[\phi_L^{-}(\boldsymbol{0})]^*=-\mathcal{M}_{3}\Theta[\phi_L^{-}(\boldsymbol{0})]^*,
\end{eqnarray}
which provides the following matrix
\begin{eqnarray}
\mathcal{M}_{3}(\boldsymbol{p}) =  \left(\begin{array}{cccc}
m_{11} & -(\kappa_{3}+m_{11}\cot(\theta/2))e^{-i\phi} & 0 & 0 \\ 
(\kappa_{3}-m_{22}\tan(\theta/2))e^{i\phi} & m_{22} & 0 & 0 \\ 
0 & 0 & 1 & 0 \\ 
0 & 0 & 0 & 1
\end{array}  \right),
\end{eqnarray}
and the associated spinor is
\begin{eqnarray}\label{psi3}
\psi_{3\downarrow}(\boldsymbol{p})=\sqrt{m}\left(\begin{array}{c}
i\kappa_{3}\mathcal{B}^{+}\boldsymbol(p)\sin(\theta/2)e^{-i\phi/2}\\
-i\kappa_{3}\mathcal{B}^{+}\boldsymbol(p)\cos(\theta/2)e^{i\phi/2}\\
-\mathcal{B}^{-}\boldsymbol(p)\sin(\theta/2)e^{-i\phi/2}\\
\mathcal{B}^{-}\boldsymbol(p)\cos(\theta/2)e^{i\phi/2}
\end{array}
\right).
\end{eqnarray}
And the last mapping possibility is given by
\begin{eqnarray}
\vec{\sigma}.\hat{p}\;\mathcal{M}_{4}\phi_L^{-}(\boldsymbol{0})=+\mathcal{M}_{4}\phi_L^{-}(\boldsymbol{0}),
\end{eqnarray}
with
\begin{eqnarray}
\mathcal{M}_{4}(\boldsymbol{p}) =  \left(\begin{array}{cccc}
1 & 0 & 0 & 0 \\ 
0 & 1 & 0 & 0 \\ 
0 & 0 & m_{33} & (\kappa_{4}+m_{33}\tan(\theta/2))e^{-i\phi} \\ 
0 & 0 & (-\kappa_{4}+m_{44}\cot(\theta/2))e^{i\phi} & m_{44}
\end{array}  \right).
\end{eqnarray}
Then, finally we obtain
\begin{eqnarray}\label{psi4}
\psi_{4\uparrow}(\boldsymbol{p})=\sqrt{m}\left(\begin{array}{c}
-i\mathcal{B}^{+}\boldsymbol(p)\cos(\theta/2)e^{-i\phi/2}\\
-i\mathcal{B}^{+}\boldsymbol(p)\sin(\theta/2)e^{i\phi/2}\\
\kappa_{4}\mathcal{B}^{-}\boldsymbol(p)\cos(\theta/2)e^{-i\phi/2}\\
\kappa_{4}\mathcal{B}^{-}\boldsymbol(p)\sin(\theta/2)e^{i\phi/2}
\end{array}
\right).
\end{eqnarray} 
Indeed $\mathcal{M}_{i}^{-1}$ exists, where $i=1,2,3,4$, ensuring an invertible map and $\mathcal{M}^{2}\neq \mathbbm{1}$. Note that for all the above spinors the representation space is connected by the identity matrix (less than a phase factor) and not by the parity symmetry or Wigner time-reversal operator. This is what leads us to state that the $\psi_{i}(\boldsymbol{p})$ spinors does not, necessarily, fulfil the Dirac dynamics.

\section{Classifying the $\psi$ spinors}\label{sec4}
Consider the Minkowski spacetime $(M,\eta_{\mu\nu})$ and its tangent bundle $
TM$. Denoting sections of the exterior bundle by
$\sec\Lambda (TM)$,  and given a $k$-vector $a \in \sec\Lambda^k(TM)$,  the {reversion} is defined by $\tilde{a}=(-1)^{|k/2|}a$, and the grade
involution by $\hat{a}=(-1)^{k}a$, where $|k|$ stands for the integral part of $k$. By extending the Minkowski metric  from $\sec\Lambda^1(TM)=\sec T^*M$ to
$\sec\Lambda(TM)$, and  considering  $a_1,a_2 \in \sec \Lambda(V)$, the
{left contraction} is given by ${g}(a \lrcorner a_1,a_2)={g}(a_1
,\tilde{a}\wedge a_2 ). $
The well-known Clifford product between the dual of a vector field $ v \in \sec\Lambda^1(TM)$ and a multivector is given by $ v a = v \wedge a+ v a $, defining thus the spacetime Clifford algebra $C\ell_{1,3}$.  The set $\{{e}_{\mu }\}$ represents sections of the frame bundle
$\mathbf{P}_{\mathrm{SO}_{1,3}^{e}}(M)$ and  $\{\gamma^{\mu }\}$ can be further thought as being  the dual
basis $\{{e}_{\mu }\}$, namely, $\gamma^{\mu }({e}_{\mu })=\delta^\mu_{\;\nu}$.
 Classical spinors are objects of the space that carries the usual
$\tau=(1/2,0)\oplus (0,1/2)$ representation of the Lorentz group, that  can be thought as being sections of the vector bundle
$\mathbf{P}_{\mathrm{Spin}_{1,3}^{e}}(M)\times _{\tau }\mathbb{C}^{4}$.

Given a spinor field $\psi \in \sec \mathbf{P}_{\mathrm{Spin}_{1,3}^{e}}(M)\times_{\tau
}\mathbb{C}^{4}$, the bilinear covariants  are sections of the bundle $\Lambda(TM)$ \cite{lounesto2001clifford, crawford1}. Indeed, the well-known Lounesto spinor classification is based upon bilinear covariants and the underlying multivector structure. The physical nature of the classification focuses on the bilinear covariants, which are physical observables, characterizing different types of fermionic particles. The observable quantities are given by the following multivectors:
\begin{eqnarray}
\label{cova}
\sigma&=&\psi^{\dag}\gamma_{0}\psi, \hspace{1.9cm}  \omega=-\psi^{\dag}\gamma_{0}\gamma_{0123}\psi,\nonumber\\ \boldsymbol{J}&=&\psi^{\dag}\mathrm{\gamma_{0}}\gamma_{\mu}\psi\gamma^{\mu},  \hspace{1cm}
\boldsymbol{K}=\psi^{\dag}\mathrm{\gamma_{0}}\textit{i}\mathrm{\gamma_{0123}}\gamma_{\mu}\psi\gamma^{\mu},\nonumber\\ \boldsymbol{S}&=&\frac{1}{2}\psi^{\dag}\mathrm{\gamma_{0}}\textit{i}\gamma_{\mu\nu}\psi\gamma^{\mu}\wedge\gamma^{\nu},
\end{eqnarray}
where $\gamma_{0123}:=\gamma_5=i\gamma_0\gamma_1\gamma_2\gamma_3$. The set $\{\mathbbm{1},\gamma
_{I}\}$ (where $I\in\{\mu, \mu\nu, \mu\nu\rho, {5}\}$ is a composed index) is a basis for
${\cal{M}}(4,\mathbb{C})$ satisfying  $\gamma_{\mu }\gamma _{\nu
}+\gamma _{\nu }\gamma_{\mu }=2\eta_{\mu \nu }\mathbbm{1}$.

The above bilinear covariants in the Dirac theory are interpreted respectively  as the  mass of the particle ($\sigma$), the pseudo-scalar ($\omega$) relevant for parity-coupling, the probability current ($\mathbf{J}$), the direction of the electron spin ($\mathbf{K}$), and the probability density of the intrinsic electromagnetic moment ($\mathbf{S}$) associated to the electron. The most important bilinear covariant for the our goal here  is $\mathbf{J}$, although with a different meaning.
A prominent requirement for the Lounesto spinor classification is that the bilinear covariants satisfies quadratic algebraic relations, namely, the so-called Fierz-Pauli-Kofink (FPK) identities
\begin{eqnarray}\label{FPKID}
J_{\mu}J^{\mu}=\sigma+\omega,\;\;\;\;J_{\mu}J^{\mu}=-K_{\mu}K^{\mu},\;\;\;\;J_{\mu}K_{\mu}=0,\;\;\;\; J_{\mu}K_{\nu}-J_{\nu}K_{\mu}=-\omega S_{\mu\nu}-\frac{\sigma}{2}\epsilon_{\alpha\beta\mu\nu}S^{\alpha\beta}.
\end{eqnarray}
The above identities are fundamental not only for classification, but also to further assert the inversion theorem\cite{crawford1}. Within the Lounesto classification scheme, a non vanishing $\mathbf{J}$ is crucial, since it enables to define the so called boomerang \cite{lounesto2001clifford} which has an ample geometrical meaning to assert that there are precisely six different classes of spinors. This is a prominent consequence of the definition of a boomerang. As far as the boomerang is concerned, it is not possible to exhibit more than six types of spinors, according to the bilinear covariants. Indeed, Lounesto spinor classification splits regular and singular spinors. The regular spinors are those which have at least one of the bilinear covariants $\sigma$ and $\omega$ non-null. Singular spinors, on the other hand, have $\sigma = 0 =\omega$, consequently the Fierz identities are normaly replaced by the more general conditions \cite{crawford1, crawford2}
\begin{eqnarray}\label{multi}
Z^{2}=4\sigma Z,\;\;\;\;Z\gamma_{\mu}Z=4J_{\mu}Z,\;\;\;\; Zi\gamma_{5}=4\omega Z,\nonumber\\
Zi\gamma_{\mu}\gamma_{\nu}Z=4S_{\mu\nu}Z,\;\;\;\;Z\gamma_{5}\gamma_{\mu}Z=4K_{\mu}Z.
\end{eqnarray}
When an arbitrary spinor $\xi$ satisfies $\xi^{*}$ and belongs to $\mathbb{C}\otimes \mathcal{C}\ell_{1,3}$ — or equivalently when $\xi^{\dagger}\gamma_{0}\psi\neq 0 \in \mathcal{M}(4, \mathbb{C})$— it is possible to recover the original spinor $\psi$ from its aggregate $Z$ given by
\begin{eqnarray}
Z=\sigma+\mathbf{J}+i\mathbf{S}+\mathbf{K}\gamma_{5}-i\omega\gamma_{5},
\end{eqnarray}
and the spinor $\xi$ by the so-called Takahashi algorithm likewise. 

Altogether, the algebraic constraints reduce the possibilities to six different spinor classes, namely
\begin{enumerate}
\item $\sigma\neq0,\;\;\;\;\;\omega\neq0.$
\item $\sigma\neq0, \;\;\;\;\;\omega=0.$
\item $\sigma=0,\;\;\;\;\;\omega\neq0.$
\item $\sigma=0=\omega,\;\;\;\;\;K_{\mu}\neq0,\;\;\;\;\;S_{\mu\nu}\neq0.$
\item $\sigma=0=\omega,\;\;\;\;\;K_{\mu}=0,\;\;\;\;\;S_{\mu\nu}\neq0.$
\item $\sigma=0=\omega,\;\;\;\;\;K_{\mu}\neq0,\;\;\;\;\;S_{\mu\nu}=0.$
\end{enumerate}
The spinors types-(1), (2) and (3), are called Dirac spinor fields (regular spinors). The spinor field (4) is called flag-dipole \cite{esk}, while the spinor field (5) is named flag-pole\cite{benn, jcap, co9}. Majorana spinors are elements of the flag-pole class. Finally, the type (6) dipole spinors are exemplified by Weyl spinors. Note that there are only six different spinor fields. For the regular case, since $\mathbf{J} \neq 0$, it follows that $\mathbf{S} \neq 0$ and $\mathbf{K} \neq 0$ as results of the identities \eqref{FPKID}. On the other hand, for the singular case, the geometry determines that $\mathbf{J}(s+h\gamma_{0123}) = \mathbf{S}+ \mathbf{K}\gamma_{0213}$. As we can see, it is important that $ \mathbf{J}\neq 0$ to ensure that FPK identities are satisfied, so we find only six different classes of spinors. In fact, a non vanishing $\mathbf{J}$ is indispensable only for the regular spinor case. As mentioned, the above classification makes use of this constraint in all the cases, since the very idea of the classification was to categorize spinors that could be related to Dirac particles in some aspect. If we ignore this physical concept, more spinors can be found.

The type-4 spinors that we find in this work, does not necessarily guarantee the existence of any physical quantum field operator. By physical, we mean that such field must be local, provide Fermi statistic and also ensure a positive-definite Hamiltonian, as firstly observed in \cite{chengflagdipole}.
After we have accomplished the mapping task, we are able to classify the spinor presented in Section \ref{sec3}.
Using as an example, the spinor given in equation \eqref{psi1}, and defining its dual structure as the same as the Dirac dual, we have 
\begin{eqnarray*}
\bar{\psi}_{1\uparrow}(\boldsymbol{p})=\sqrt{m}\left(\begin{array}{cccc}
\mathcal{B}^{-}(p)\cos(\theta/2)e^{i\phi/2} & \mathcal{B}^{-}(p)\sin(\theta/2)e^{-i\phi/2} & -i\kappa_{1}^{*}\mathcal{B}^{+}(p)\cos(\theta/2)e^{i\phi/2} & -i\kappa_{1}^{*}\mathcal{B}^{+}(p)\sin(\theta/2)e^{-i\phi/2}
\end{array}
\right).
\end{eqnarray*}
The next step is to calculate the 16 bilinear covariants \cite{lounesto2001clifford}, giving us to the following physical observables
\begin{eqnarray*}
\sigma&=&im(\kappa_{1}-\kappa^{*}_{1}),\\
\omega&=&m(\kappa^{*}_{1}-\kappa_{1}),
\end{eqnarray*}
\begin{eqnarray*}
J_{0}&=&(|\kappa_{1}|^{2}-1)E+(|\kappa_{1}|^{2}+1)p,\\
J_{1}&=&\bigg((1-|\kappa_{1}|^{2})E-(|\kappa_{1}|^{2}+1)p\bigg)\sin\theta\cos\phi,\\
J_{2}&=&\bigg((1-|\kappa_{1}|^{2})E-(|\kappa_{1}|^{2}+1)p\bigg)\sin\theta\sin\phi,\\
J_{3}&=&\bigg((1-|\kappa_{1}|^{2})E-(|\kappa_{1}|^{2}+1)p\bigg)\cos\theta,
\end{eqnarray*}
\begin{eqnarray*}
K_{0}&=& 2p,\\
K_{1}&=&-\bigg[\big(1+|\kappa_{1}|^{2})E+(|\kappa_{1}|^{2}-1)p\bigg],\\
K_{2}&=&-\bigg[\big(1+|\kappa_{1}|^{2})E+(|\kappa_{1}|^{2}-1)p\bigg]\sin\theta\sin\phi,\\
K_{3}&=&-\bigg[\big(1+|\kappa_{1}|^{2})E+(|\kappa_{1}|^{2}-1)p\bigg]\cos\theta,
\end{eqnarray*}
\begin{eqnarray*}
S_{01}&=&\frac{m}{2}\bigg[\kappa_{1}+\kappa_{1}^{*}\bigg]\sin\theta\cos\phi,\\
S_{02}&=&\frac{m}{2}\bigg[\kappa_{1}+\kappa^{*}_{1}\bigg]\sin\theta\sin\phi,\\
S_{03}&=&\frac{m}{2}\bigg[\kappa_{1}+\kappa^{*}_{1}\bigg]\cos\theta,\\
S_{12}&=&\frac{im}{2}\bigg[\kappa_{1}-\kappa^{*}_{1}\bigg]\cos\theta,\\
S_{13}&=&\frac{-im}{2}\bigg[\kappa_{1}-\kappa^{*}_{1}\bigg]\sin\theta\sin\phi,\\
S_{23}&=&\frac{im}{2}\bigg[\kappa_{1}-\kappa^{*}_{1}\bigg]\sin\theta\cos\phi.
\end{eqnarray*}
Accordingly, the resulting spinor is classified as regular spinor when $\kappa_{1}\in\mathbb{C}$, in agreement with \cite{diracpauli} or type-4 when $\kappa_{1}= 1 $. Note that it is the only possible real value, if one imposes any other real value for $\kappa$, it can be shown that the FPK identities, given by \eqref{multi}, will not be fulfilled. So, in this specific framework, we had an unexpected theoretical discovery: type-4 spinors naturally arised from a mapping process between two different classes of spinor.
Here we choose to abstain from discussions about using a complex phase parameter due to the reasons that any relevant information about this case can be found in Ref \cite{diracpauli}.

As showed in reference \cite{bilineares}, Elko mass-dimension-one spinors do not belong to the Lounesto classification due to the fact that they are endowed with different dual structure and dual-helicity features, a specific classification for these spinors remains open. In this communication we constructed the first mass-dimension-one fermion which carry single-helicity feature. We have developed a connection between the Lounesto classification and a special spinor type within the mass-dimension-one classification, it is the only known fermion that compose this new classification. If more mass-dimension-one fermions (like Elko) exist, another mapping treatment can be developed and perhaps a general, robust and complete connection betwen Lounesto and MDO classification could be established.

\section{Type-4 spinors Underlying Features: Dynamic and Spin sums}\label{sec5}
As mentioned before, type-4 spinors do not satisfy the Dirac equation as previosly demonstrated by Lounesto \cite{lounestolivro}. Thus, we concluded that they are not eigenspinors of the parity operator ($P=m^{-1}\gamma_{\mu}p^{\mu}$), they  also do not belong to a set of eigenspinors of the charge conjugation operator due to the fact that their representation spaces are not connected by the Wigner time-reversal operator ($\Theta$). Moreover, as a characteristic inherited from Elko spinors, type-4 spinors only satisfy the Klein-Gordon wave equation. 

Thus, the spin sum is given by\footnote{Please note that here we have fixed $\kappa_1=\kappa_2= \kappa_3=\kappa_4=1$, in agreement with the discussions made in Sec \ref{sec4}.}
\begin{equation}\label{spinsum}
\sum_{i=1}^{4}\psi_{i}(\boldsymbol{p})\bar{\psi}_{i}(\boldsymbol{p})=2mS(p),
\end{equation}
where the $S(p)$ matrix is defined as
\begin{equation}
S(p)=\left(\begin{array}{cccc}
0 & 0 & \frac{E+p}{m} & 0 \\ 
0 & 0 & 0 & \frac{E+p}{m} \\ 
\frac{E-p}{m} & 0 & 0 & 0 \\ 
0 & \frac{E-p}{m} & 0 & 0
\end{array}  \right),
\end{equation}
or, in other words, 
\begin{equation}\label{dinamica666}
S(p)= \frac{1}{m}(E\gamma_0 - p\gamma_0\gamma_5),
\end{equation}
where $\gamma_5=-i\gamma_0\gamma_1\gamma_2\gamma_3$. Note that $S(p)$ is not Lorentz invariant.
The $S(p)$ operator present the following properties $S^{2}(p)=\mathbbm{1}$ and $S^{-1}(p)$ exist.
An interesting fact that must be stressed is that
\begin{equation}\label{dinamica1}
[S^2(p)-\mathbbm{1}]\psi_{i}(\boldsymbol{p}) = 0, 
\end{equation} 
such that, the operator that appears on the right hand side in equation \eqref{spinsum} annihilates type-4 spinors. At this stage we are looking for all the rudimentary mathematical and physical details encoded on type-4 spinors. Once they are well established, we propose the field quantization where type-4 spinors play a role of expansion coefficients of the quantum field. 

Note that it is also possible to redefine the dual structure of the above spinors, ensuring a Lorentz invariant spin sum and providing new physical information. Such redefinition give allowance for a Lorentz invariant theory but the new bilinear structure do not respect FPK identities so we have chosen to abandon the dual redefinition.

\section{Final Remarks}\label{sec6}

The present paper reports the discovery of type-4 spinors. We constructed the first case of mass-dimension-one fermions endowed with single helicity. Our theoretical results create a new class of single-helicity spinors, that are not eigenspinor of charge conjugation, parity or time-reversal symmetry. 
In additon, we have shown that the type-4 spinors only carry relevant physical information and satisfy the FPK identities if we chooses to set $\kappa=1$, otherwise, we can not guarantee it. We also showed that type-4 spinors do not fulfil the Dirac dynamic and found a non-conventional dynamics, which is given on the right hand side of the spin sums presented in \eqref{spinsum} and \eqref{dinamica666}.

It is also possible to redefine the dual structure which lead us to an invariant and non-vanishing norm under the orthonormal relation. This reveals new dynamics and new physical observables codified in this structure. Nevertheless, the price to be paid is that the related bilinear structures do not respect the FPK identities and should be deformed as was shown in \cite{bilineares}. In this scenario, these spinors would not belong to the Lounesto classification anymore.

From the physical point of view we highlight that this mapping procedure shows that once we break the chirality symmetry of the Elko spinor we are breaking the link between the representation spaces of the resulting spinors. Therefore, these new spinors only inherit some characteristics of the originating spinor, and, in addition, such symmetry break brings a breach between the representation spaces, making the resulting spinors even more exotic.

\section{Acknowledgements}
The authors express their gratitude to Prof. Julio Marny Hoff da Silva and Prof. Jos\'e Abdalla Helay\"el-Neto for careful reading the entire first draft of the manuscript, providing many insightful suggestions and questions. We also thanks to Dr. Oswaldo Miranda for fruitful discussions in the final stage of the work. 
RJBR thanks CAPES and CNPq (Grant Number 155675/2018-4) for the financial support and CHCV thanks CNPq (PCI Grant Number 300381/2018-2) for the financial support.

\bibliographystyle{unsrt}
\bibliography{refs}

\end{document}